\documentclass[conference]{IEEEtran}
\IEEEoverridecommandlockouts
\usepackage{cite}
\usepackage{amsmath,amssymb,amsfonts,amsthm, latexsym}
\usepackage{graphicx}
 \graphicspath{{./img/}}

\usepackage[hyphens]{url}
\usepackage[english]{babel}
\usepackage[autostyle]{csquotes}
\MakeOuterQuote{"}

\usepackage[font=scriptsize]{caption}
\usepackage{float}
\addto\captionsenglish{\renewcommand{\figurename}{Fig.}}

\usepackage{verbatimbox}

\begin{document}

\title{Modelling Technique for GDPR-compliance: Toward a Comprehensive Solution}

\author{\IEEEauthorblockN{Naila Azam\IEEEauthorrefmark{1},
Anna Lito Michala\IEEEauthorrefmark{1},
Shuja Ansari\IEEEauthorrefmark{2},
Nguyen B. Truong\IEEEauthorrefmark{1}
}
\IEEEauthorblockA{\IEEEauthorrefmark{1}
School of Computing Science, University of Glasgow, United Kingdom \\
}
\IEEEauthorblockA{\IEEEauthorrefmark{2}
James Watt School of Engineering, University of Glasgow, United Kingdom \\
Emails: n.naila.1@research.gla.ac.uk\}\{Annalito.Michala, Shuja.Ansari, and Nguyen.Truong\}@glasgow.ac.uk}
}

\maketitle

\begin{abstract}
Data-driven applications and services have been increasingly deployed in all aspects of life including healthcare and medical services in which a huge amount of personal data is collected, aggregated, and processed in a centralised server from various sources. As a consequence, preserving the data privacy and security of these applications is of paramount importance. Since May 2018, the new data protection legislation in the EU/UK, namely the General Data Protection Regulation (GDPR), has come into force and this has called for a critical need for modelling compliance with the GDPR's sophisticated requirements. Existing threat modelling techniques are not designed to model GDPR compliance, particularly in a complex system where personal data is collected, processed, manipulated, and shared with third parties. In this paper, we present a novel comprehensive solution for developing a threat modelling technique to address threats of non-compliance and mitigate them by taking GDPR requirements as the baseline and combining them with the existing security and privacy modelling techniques (i.e., \textit{STRIDE} and \textit{LINDDUN}, respectively). For this purpose, we propose a new data flow diagram integrated with the GDPR principles, develop a knowledge base for the non-compliance threats, and leverage an inference engine for reasoning the GDPR non-compliance threats over the knowledge base. Finally, we demonstrate our solution for threats of non-compliance with legal basis and accountability in a telehealth system to show the feasibility and effectiveness of the proposed solution.
\end{abstract}

% Note that keywords are not normally used for peer-review papers.

\IEEEpeerreviewmaketitle

\section{Introduction}
In this digital era, personal data is frequently being processed and shared with various data-driven applications such as finance, automotive, and health services. The wide use of those data-hungry applications puts our data at high risk of data and privacy breaches. Numerous threat modelling techniques such as STRIDE and LINDDUN are utilised when designing the applications. However, these techniques are either insufficient or not suitable to model the threats of non-compliance with sophisticated data protection regulations such as the General Data Protection Regulation (GDPR) \cite{voigt2017eu} in the EU/UK \cite{azam2022data}.

The existing modelling techniques focus on pre-defined software-based security threats and, consequently, are limited to modelling privacy attacks or any new types of threats including non-compliance in complex data-driven systems \cite{azam2022data}. Compliance with the GDPR requires the protection of individuals' data rights and privacy while expanding the duties and responsibilities of data controllers, as a result, it is significantly more effective at ensuring users' privacy and protecting personal data. However, traditional data privacy threat models cannot completely fulfil these obligations, which only concentrate on various kinds of privacy-related attacks \cite{015}. Therefore, there is a dire need for a modelling technique to determine and mitigate the threats of GDPR non-compliance in various applications and services.

In this paper, we propose a novel GDPR-compliance modelling solution that harmonises the existing security and privacy modelling techniques (i.e., STRIDE and LINDDUN) with the GDPR baselines including entity roles, obligations, and data protection principles. The major contributions are three-fold as follows:

\begin{enumerate}
 \item A holistic solution approach for designing and developing a GDPR-compliance threat modelling technique including modules and components with specifications, requirements, and interactions among the components.
 \item Provide an implementation reference including the creation of a knowledge base for GDPR compliance, a novel Data flow Diagram (DFD), and an inference engine to generate the non-compliance threats. 
 \item The demonstration of the proposed solution for non-compliance threats with legal foundation and accountability in the Telehealth Service System (TSS) use-case.
\end{enumerate}

The rest of the paper is organised as follows: Section \ref{sec2} provides background and related work on threat modelling for security and data privacy. Section \ref{sec3} is dedicated to presenting a holistic solution approach for a GDPR-compliance modelling technique. Section \ref{sec4} showcases the implementation reference to develop the modelling technique with reference to the use-case of telehealth services. Section \ref{sec5} presents the demonstration results with an insightful discussion on the suitability, feasibility and performance of the proposed technique. Section \ref{sec6} is to summarise the work with potential research directions.

\section{Background and Related Work}\label{sec2}
This section provides background on threat modelling techniques, particularly for data privacy and GDPR compliance.

\subsection{Significant threat modelling techniques}
Existing threat modelling techniques are discussed in this section with a focus on their advantages and disadvantages.

\subsubsection{STRIDE}
Spoofing, Tempering, Repudiation, Information Disclosure, and Elevation of Privileges (STRIDE) developed by Microsoft \cite{scandariato2015descriptive} has found applications in various cyber-physical systems because of the ability to represent threats related to data flows\cite{gd94}. This modelling approach is static and exploratory at the design stage for threat mitigation techniques identification \cite{abbas2021identifying}. However, STRIDE does not scale well with complexity and lacks in modelling the accuracy of threats to data privacy \cite{gd95}.

\subsubsection{LINDDUN} 
Linkability, Identifiability, Non- Repudiation, Detectability, Information Disclosure, Unawareness, and Non-Compliance (LINDDUN) is developed based on STRIDE with the purpose of modelling data privacy threats \cite{014}. LINDDUN iteratively builds threat trees of the identified threats \cite{015} which makes the process time-consuming. Moreover, though LINDDUN has extensive privacy documentation, it is dependent on certain assumptions and lacks adaptability in complex situations where several components interact \cite{015} and face expressiveness limitations \cite{gd155}. Such limitations relating to trust and attacker capabilities perhaps should be considered for completeness \cite{gd155}. In \cite{azam2022data}, the authors have given a comparative analysis of various threat modelling methods based on the criteria of maturity, focus, time/effort, mitigation etc. and shown that regulatory compliance on data privacy is not inherent in these approaches. 

\subsection{General Data Protection Regulation and Compliance}
The General Data Protection Regulation (GDPR) is the legislation regulating data security and privacy in the EU since May 2018 \cite{gd121}. GDPR enforces the protection of confidentiality of personal data by adhering to its principles under rigorous standards. Non-compliance can result in significant fines \cite{gd1}. The GDPR introduces seven fundamental principles that must be followed, namely lawfulness, fairness, and transparency; data minimization; purpose limitation; storage limitation; accuracy; and accountability.

GDPR goes further to allocate responsibilities of adherence to entities; namely Data Controller (DC), Data Processor (DP), and Data Subject (DS) \cite{gd4}. The DS has the rights: to be informed, of access, rectification, erasure, restrict processing, object, and data portability, and to automated decision-making, with the DC and DP responsible for the delivery of those rights. There are six legal grounds for processing data which should be in place in advance of processing: consent, a legitimate interest, a contract, a legal need, a vital interest, and a public interest. Finally, GDPR enforces the adoption of proper controls and statistical disclosure-limitation techniques.

\subsection{Related Work} 
In light of data protection compliance threat analysis, the authors in \cite{surridge2019modelling} have proposed a framework for modelling compliance in which a System Security Modeller (SSM) tool has been developed. This tool is expected to enable the automated detection of compliance issues and end-to-end security concerns during system layout. In \cite{robol2017toward}, the authors have provided a modelling framework motivated by the \textit{privacy-by-design} concept for designing systems that are GDPR-compliant. Semantic web technologies are also leveraged to represent and query provenance data pertaining to GDPR compliance requirements \cite{pandit2018queryable}. The authors have developed a provenance ontology called GDPRov\footnote{\url{https://harshp.com/GDPRov/}} to express provenance data on consent and data lifecycles. A linked data version of the GDPR text and an ontology defining its many terminology and concepts are both provided by GDPRtEXT in the same work \cite{pandit2018gdprtext}. The semantic web-based approach enables the creation of meaningful knowledge in terms of concepts and relationships with the flexibility to be developed and connected in accordance with requirements. For instance, an interactive ontology for GDPR has been developed by Irem Besik\footnote{\url{https://github.com/irembesik/gdpr-ontology}}.

These individual efforts are either designed for a specific purpose or are not considered under the big picture of a GDPR-compliance modelling technique. The knowledge domains are scattered and not as useful as their potential. Nevertheless, such research works have paved the way for us to integrate and utilise the existing knowledge about the GDPR for the complete modelling technique, thanks to the advancement of Semantic Web technologies.

\section{GDPR-compliance threat model: A Holistic Solution Approach}\label{sec3}

\subsection{Overview of the solution approach}
\begin{figure}[ht]
     \centering
     \includegraphics[width=0.3\textwidth]{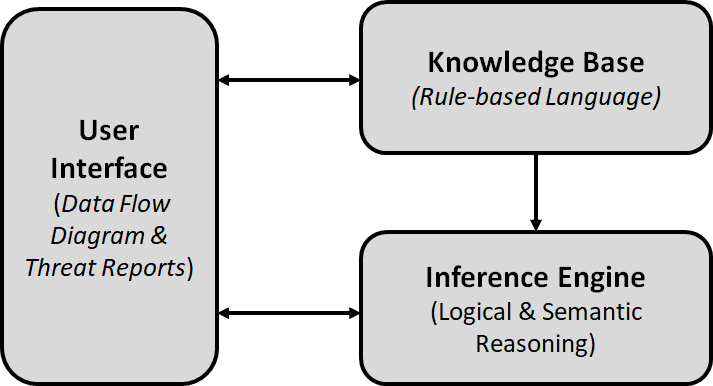}
     \caption{A high-level system architecture of a GDPR-compliance threat modelling tool}
     \label{fig:expert-system}
\end{figure}

This section provides an overview of the solution approach to design and develop a modelling technique for GDPR compliance. From our perspective, the proposed modeller is similar to an expert system for the GDPR-compliance domain \cite{voigt2017eu} which is comprised of (i) an interface to help a modeller design a specific system, (ii) a knowledge base for this system, and (iii) an inference engine to do the reasoning over the knowledge base (Fig. \ref{fig:expert-system}).

\subsection{Building a rule-based Knowledge base}
\begin{figure}[ht]
    \centering
    \includegraphics[width=0.3\textwidth]{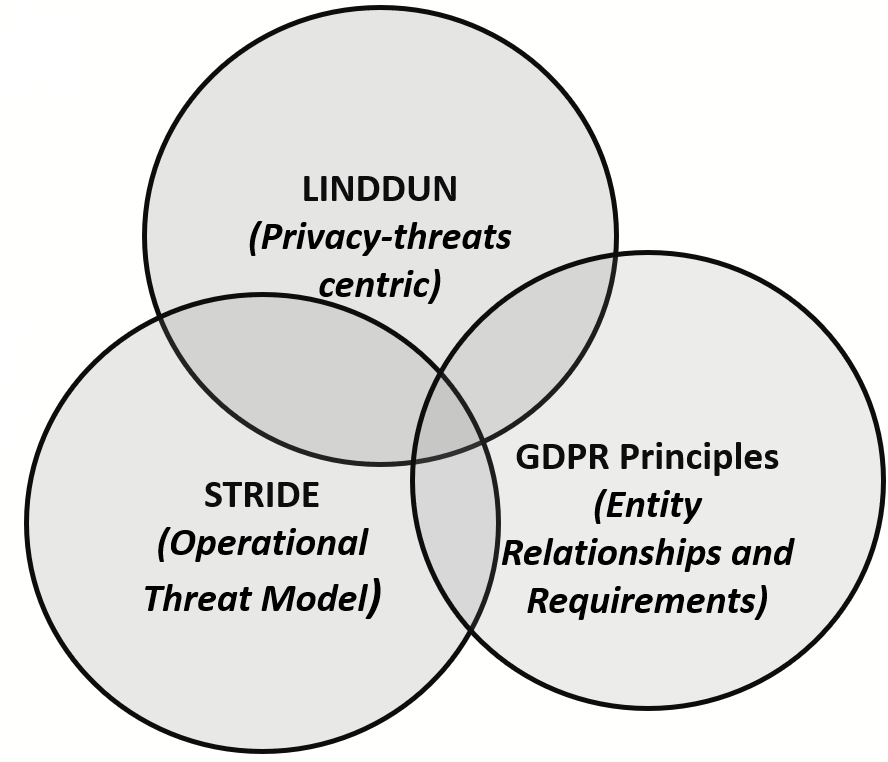}
    \caption{GDPR-Compliance Modelling Catalyst}
    \label{fig:catalyst}
\end{figure}

The knowledge base is represented using rule-based/policy-based language such as RuleML and Semantic Web Rule Language (SWRL). Every rule specifies a relation, recommendation, directive, strategy or heuristic and has the $IF(condition)$ $THEN(action)$ structure. As illustrated in Fig. \ref{fig:catalyst}, the knowledge base for the GDPR non-compliance threat modelling consists of three main areas: (i) STRIDE knowledge base: rule-based threat library obtained from STRIDE, LINDDUN knowledge base: additional threat trees described in LINDDUN specification and converted to rule-based language; and (iii) the GDPR knowledge base: obtained from Ontology and expert knowledge using an SWRL as a combination of the Web Ontology Language (OWL) and the Rule Markup Language (RuleML)\footnote{\url{https://www.w3.org/Submission/2004/SUBM-SWRL-20040521/}}.
 
The three aspects, i.e., security, data privacy, and GDPR-compliance sets of knowledge, interplay with each other to make sure whether a service provider is compliant with the GDPR. When developing the modeller, we employ STRIDE-the operational threat model with DFD along with data security threats, LINDDUN-privacy threats tree, and the GDPR-compliance baseline including the legal basis, accountability and governance, DS rights, and DS, DC, and DP relationships. The knowledge base is comprised of two parts: a default knowledge base and a system-specific knowledge base as follows.

\subsubsection{Default Knowledge base}
The default knowledge base for the GDPR non-compliance threat modelling consists of three overlapped areas: (i) the STRIDE knowledge base converted to a rule-based threat library, (ii) the LINDDUN knowledge base which is additional threat trees converted to rule-based knowledge, and (iii) the GDPR knowledge base. Again, all of this knowledge is represented under Rule-based/Policy-based language such as RuleML \cite{boley2010ruleml} and Semantic Web Rule Language (SWRL)\cite{horrocks2004swrl}.

The first two areas are extracted from the existing threat modelling tools whereas the GDPR-compliance knowledge base is developed by our team. For instance, we have taken consent as the legal basis for processing personal data and used \textit{GConsent} the \textit{OWL2 ontology} to represent consent for GDPR compliance. The ontology is based on an analysis of modelling metadata requirements related to the consent lifecycle for GDPR compliance. For example, the consent complaint requirements in our knowledge base are used as expressions such as "\textit{ConsentProvided}" and "\textit{ConsentRequestFormProvided}".

\subsubsection{System-specific Knowledge base}
The second part of our knowledge base is the specific knowledge of a particular system to be modelled. This is the duty of the modeller, who understands the system and will use an existing tool to provide such information to the knowledge base. For example, for our use-case, the Telehealth Services System (TSS), the modeller will add system-specific information to the knowledge base using a provided tool (e.g., the Microsoft Threat Modelling Tool (MSTMT)) with the novel format of the DFD.

\subsection{Developing an inference engine}
As illustrated in Fig. \ref{fig:expert-system}, an inference engine is performed over the knowledge base to reason about potential compliance threats in a system. As our knowledge base is rule-based language, a rule-based inference engine is utilised for reasoning, for instance, a semantic reasoner with either \textit{backward-chaining} or \textit{forward-chaining} algorithms. This reasoner takes the knowledge base as its input and iteratively infers new knowledge until a goal has been reached (i.e., finding a specific non-compliance threat) or no rules can be matched (i.e., finding all potential non-compliance threats). For instance, in the demonstration in Section \ref{sec5}, we use the MSTMT's inference engine which follows the following defined syntax to infer potential threats.

\begin{verbnobox}[\fontsize{9pt}{9pt}\selectfont]
Include: IF x is A, and y is B or C is r 
Exclude: IF s is F, or u is G and v is H
THEN z= px + qy * tr
\end{verbnobox}

The terms $Include$ and $Exclude$ work as $IF(condition)-THEN(conclusion)$ rules. If the condition is satisfied as defined in \textit{Include}, then whatever condition is written in $Exclude$ the inference engine should exclude it for the defined threats in the knowledge base. Where \textit{A, B, C, F, G}, and \textit{H} are the sets in the antecedent. While \textit{p, q} and \text{r} are all constants.

\subsection{A Novel Data Flow Diagram}
GDPR concepts are integrated into the legacy DFD to form a novel DFD. For this reason, GDPR-related roles are introduced to show adherence to the system's regulations. The novel DFD is an idea to complement the existing DFD defined in STRIDE with the new entities and their relationships (Fig. \ref{fig:novelDFD}).

\subsubsection{New Entities for modelling GDPR compliance}
The new entities are defined for the GDPR roles such as Data Subject (DS), Data Controller (DC), and Data Processor (DP). Other entities such as Supervisory Authority (SA) (i.e., a government entity to govern compliance with the GDPR) and Reporting Mechanism (RM) (i.e., where DS can lodge a complaint against any data breach, and DC and DP can report any data breach through RM to Supervisory Authority) are also defined. Compliance Trust Boundary Border is also implemented to present that compliance trust boundary would be where a system attains an increased privilege level of compliance. In Fig. \ref{fig:novelDFD} the circle shape is depicted as processes, the square shape reflects the external entity (e.g., DS), and the entity with a rectangle shape is the traditional Generic Data Store (GDS). Some entities with their attributes are defined below:

\begin{verbnobox}[\fontsize{9pt}{9pt}\selectfont]
Element1 Data Controller(DC)
Actions: Provide; Request; Notify; Response;
Accomplish
Properties: ConsentRequestForm; CleanData;
ErasingData; EraseDataWithin28Days; DataBreach
\end{verbnobox}
\begin{verbnobox}[\fontsize{9pt}{9pt}\selectfont]
Element2: Data Subject(DS)
Actions: Provide; Request; Complain
Properties: Consent; ErasingData;
DataBreach
\end{verbnobox}

\subsubsection{New Relationships for modelling GDPR compliance}
\begin{figure}[ht]
     \centering
     \includegraphics[width=0.45\textwidth]{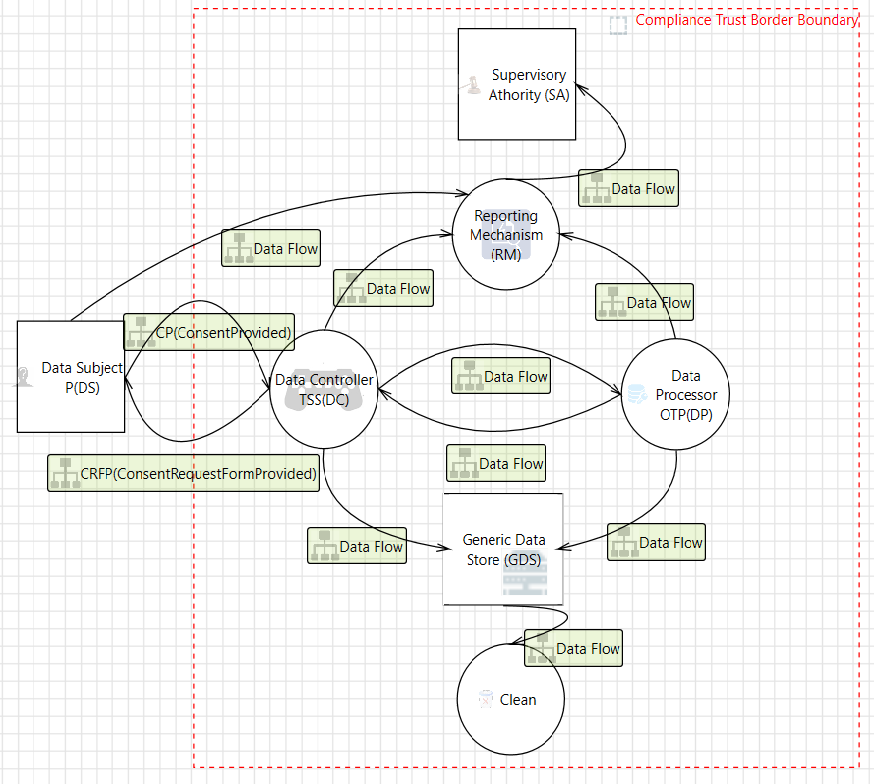}
    \caption{The proposed DFD can specify data flow between the new entities (GDPR-related) and the traditional entities (System-related)}
     \label{fig:novelDFD}
\end{figure}

The interactions among various entities in a complex system result in challenging tasks to model GDPR-compliance \cite{pandit2018gdpr}. In this project, we define a variety of interactions between the new entities with attributes in order to help the modellers specify their system in detail. In our demonstration, we build new types of relationships between entities in the proposed DFD with attributes based on the GDPR requirements using the MSTMT. For instance, as shown in Fig. \ref{fig:novelDFD} the DS-DC relationship will have some attributes such as $ConsentProvided$, and $RequestForErasingData$ with detailed information on the $Consent$ and \textit{Right to Erasure} as follows:

\begin{verbnobox}[\fontsize{9pt}{9pt}\selectfont]
Relationship1: DS-DC, DC-DS
Properties: (DS-DC)ConsentProvided, 
RequestForErasingData;
(DC-DS)ConsentRequestFormProvided
\end{verbnobox}

Modellers then can select the relationships with attributes in their systems, providing specific knowledge for the inference engine to accurately determine potential non-compliance threats (i.e., non-consent and non-provided right to erasure).

\section{Use-case: Modelling the GDPR compliance for Tele-health services System}\label{sec4}
This section describes how the threat modelling technique is employed in Telehealth services System (TSS) \cite{abomhara2015stride}.

\subsection{Consent and Right to Erasure in Telehealth services}
Telehealth is the application for the provision of a variety of user-group-specific healthcare services to individuals (e.g., patients, physicians, nurses, etc.) who are located in a diverse range of locations \cite{sosa1997internet}. The TSS has been facing various challenges related to the security and privacy of patients. For example, inadequate security procedures allow for potential data breaches \cite{abomhara2015stride}, causing patients and healthcare professionals to be vulnerable to security and privacy threats \cite{vashist2014commercial}. STRIDE was employed to identify every possible security threat to TSS in \cite{abomhara2015stride} and it has shown that the non-compliance threats could result from STRIDE security threats emerging within TSS (e.g., non-consent, non-providing right to erasure, and non-accountability).

In TSS, regardless of any legitimate interest, patients should be requested for consent to process data. The patients might also be requested for consent before their data is processed or shared with third parties by a data processor (such as an Organ Transplant Service). Moreover, a patient may request to erase his/her personal data from the data stores; and the right to erasure may be violated if the DC or DP fails to provide the DS with the required data erasure, posing a non-compliance threat of the non-provided right to erasure.

\subsection{Data Flow Diagram for Telehealth Services}
The proposed novel DFD is mapped on the roles of telehealth services as shown in Fig. \ref{fig:novelDFD}. As Patient (P) plays the role of the DS; Telehealth Service Server takes the role of DC; and Organ Transplant Service (OTS) is the DP in the system. Therefore, these entities are playing two separate tasks for the GDPR-related roles and the system-related roles.

As STRIDE does not provide a mechanism for modellers to add GDPR-related information to the DFD, the ultimate purpose of developing the novel DFD is for the modellers to describe their systems in relation to the GDPR legislation. In our proposed modelling tool, modellers can add a variety of entities, relationships, and events with associated characteristics so that it can help infer how the system demonstrates GDPR compliance.

For example, in Fig. \ref{fig:novelDFD}, the relationship between $P(DS)$ and $TSS(DC)$ with the annotations of $CP(ConsentProvided)$ and $CRFP(ConsentRequestFormProvided)$ reflects that the $P(DS)$ provides consent when the consent request form is provided by $TSS(DC)$ for processing the personal data with all of its compliance requirements (i.e., specific, cleared, and direct etc.). The tool automatically determines that consent has been granted and does not add a non-consent threat to the list of threats in the report. Additionally, the $clean$ process is introduced in the DFD to reflect that the DS would be able to exercise \textit{Right to Erasure} so that personal data is completely erased from the GDS. However, neither the request for \textit{Right to Erasure} from DS is illustrated, nor are its compliance requirements met by DC or DP. As a result, the MSTMT tool equipped with the proposed DFD would generate the threat of a non-provided \textit{Right to Erasure}.
 
\subsection{Non-compliance threats}
The non-compliance threats (i.e., non-consent, and non-provided \textit{Right to Erasure}) that might occur in the TSS use case  are presented in detail as follows:

\begin{verbnobox}[\fontsize{9pt}{9pt}\selectfont]
Threat type: non-consent
IF DS.Provide{Consent}=NOT AND
DC.Provide{DS.ConsentRequestForm}=NOT
THEN {non-Consent}
\end{verbnobox}

\begin{verbnobox}[\fontsize{9pt}{9pt}\selectfont]
Threat type: non-provided right to erasure
IF DS.Request{DC.EraseData} AND 
DC.Request{GDS.CleanData}=NOT AND
DC.Request{DP.EraseData}=NOT AND
DP.Request{GDS.CleanData}=NOT OR
GDS.Response.{cleanData}=Not AND
DC.Notify{RecipientAboutErasingData}=NOT AND
DP.Notify{RecipientAboutErasingData}=NOT AND
DC.Accom Request{EraseDataWithin28Days}=NOT AND
DP.Accom Request{EraseDataWithin28Days}=NOT    
THEN {non-provided right to erasure}
\end{verbnobox}

A non-compliance threat might be a consequence of more than one threat in different categories inferred by STRIDE. For instance, the threat of \textit{non-Accountability} can be an aftermath of some types of security threats identified by STRIDE itself. However, our technique shows that even a system without any security threats related to the \textit{non-Accountability} principle, still, if the system does not implement an RM for which DC or DP can notify SA of data breaches then there will be the threat of non-accountability.

In the telehealth use case, the RM process (depicted from the DFD) demonstrates that the $TSS(DC)$ and $OTS(DP)$ are responsible for notifying SA of data breaches. Otherwise, there would be a threat of non-accountability caused by either $TSS(DC)$ or $OTS(DP)$. The rules for identifying non-accountability threat is depicted below:

\begin{verbnobox}[\fontsize{9pt}{9pt}\selectfont]
Threat type: non-accountability

IF DS.Complain{RM.DataBreach} AND
DC.Report {RM.DataBreach}=NOT AND
DP.Report{RM.DataBreach}=NOT

THEN {non-accountability}
\end{verbnobox}

\subsection{Templates for Modelling GDPR-compliance Threats}
A new template for our proposed modelling technique is developed in addition to the built-in templates developed by MSTMT. This template for the GDPR-compliance threat modelling is designed to implement all of the new entities we have introduced (i.e., GDPR role-based entities and relationships) along with pre-defined rules associated with the entities. As a result, this template supports modellers to understand more about the GDPR-compliance requirements and easily model their systems using the tool.

To demonstrate the compliance threats for the use case, a new template designed for TSS has been developed (along with a DFD for the use case illustrated in Fig. \ref{fig:novelDFD}). The DFD uses this template to model the TSS system in an effective and convenient manner. The template is also available on GitHub for reference\footnote{\url{https://github.com/nailaazam/ModellingGDPRCompliance/blob/main/template/nonCompliant.tb7}}.

We have showcased how the proposed modelling technique for non-compliance threats can be employed for the TSS. The next section will further discuss and analyse the results.

\section{Results, Analysis and Discussion} \label{sec5}
In this section, results from our proposed technique for the TSS use case are presented. An insightful analysis and discussion of the results are also provided.

\subsection{GDPR-compliance Threat Reports}
\begin{figure}[ht]
     \centering
     \includegraphics[width=0.48\textwidth]{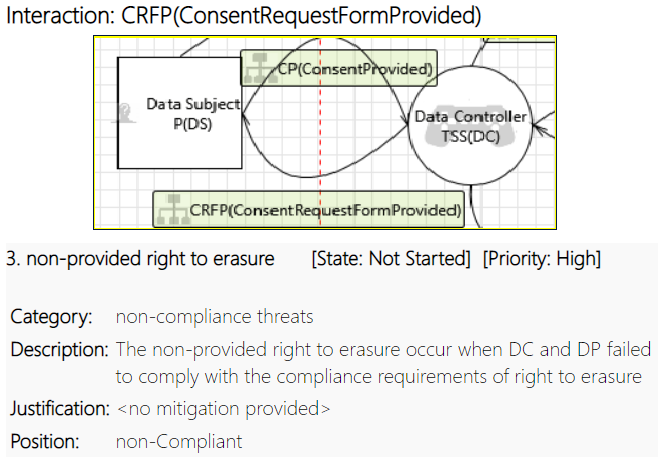}
     \caption{A part of the GDPR-compliance threats report generated by MSTMT}
     \label{fig:template}
\end{figure}

For the demonstration, we use MSTMT equipped with the proposed DFD to identify GDPR-compliance potential threats for the TSS use case. The MSTMT with the novel DFD provides a facility for modellers to describe their system in regard to the GDPR legislation, which can then be further translated into the knowledge base on the back end (i.e., \textit{System-specific Knowledge base}). Combining with the \textit{Default Knowledge base}, the list of potential threats is generated by sparking the MSTMT built-in inference engine over the whole knowledge base (i.e., clicking on the \textit{'Generate a Report'} button in MSTMT). Based on the output from the inference engine, a report is generated which shows a list of potential non-compliance threats\footnote{\url{https://github.com/nailaazam/ModellingGDPRCompliance}}. Fig. \ref{fig:template} shows a part of the non-compliance threat report that we have obtained from the tool.

Threats to non-provided \textit{Rights to Erasure} and \textit{non-Accountability} are identified across the TSS entities and recorded in the report results. Even though the DFD does not provide information to fulfil the compliance requirements of the \textit{Right to Erasure} and \textit{Accountability} principle, the report does not result in a \textit{non-Consent} threat. This is also because in DFD there is an illustration of consent requirements in the form of expressions (i.e., $CP(ConsentProvided)$ and $CRFP(ConsentRequestFormProvided)$). 

\subsection{Analysis and Discussion}
We have demonstrated the three potential non-compliance threats i.e., non-consent, non-provided right to erasure, and non-accountability for the TSS use case. As shown in Table \ref{table:threats}, symbol (×) shows the mapping between non-compliance threats and the entities (i.e., the source of the threats) defined in the DFD. As already mentioned the report does not generate the \textit{non-Consent} threat as the system described in the DFD is compliant with the legal base (i.e., Consent) requirement. However, the non-provided \textit{Right to Erasure} and \textit{non-Accountability} threats occur across all of the entities $TSS(DC)$, $OTP(DP)$, $GDS$ except the $P(DS)$. This is because there is not enough information related to the compliance requirements that can be obtained from the DFD for further analysis of potential non-compliance threats.

\begin{table}[ht]
\scriptsize
 \caption{Mapping between GDPR-compliance Threats and Sources of Threats}
\centering
\begin{tabular}{|p{2.5cm}|p{1cm}|p{1cm}|p{0.8cm}|p{0.8cm}|}
\hline
\textbf{Compliance Threats}
& \textbf{TSS(DC)} & \multicolumn{1}{l|}{OTP(DP)} & \multicolumn{1}{l|}{GDS} & \multicolumn{1}{l|}{P(DS)}  \\ \hline
\textbf {Consent} & \centering  & \centering  & \centering  & 
\\ \hline
\textbf {Right to Erasure} & \centering ×  & \centering × & \centering ×
&  
\\\hline
\textbf  {Accountability} & \centering × & \centering × &    &                   
\\ \hline
\end{tabular}
\label{table:threats}
\end{table}

The results show that if necessary compliance requirements can be obtained from the DFD then the related non-compliance threat will not be presented in the report. It can be understood that the system, as illustrated by the DFD, is compliant with this type of GDPR requirements, except for providing more information to describe the system in more detail. On the other hand, we have shown that a non-compliance threat (i.e., non-accountability) can still occur in a system even though there is no security threat related to it. These results of identifying non-compliance threats from a variety of information sets are evidence of the feasibility and effectiveness of the proposed GDPR-compliance threat modelling technique. 
\section{Conclusion and Future Work}\label{sec6}

We have proposed a holistic solution for a threat modelling technique to address the issues of non-compliance and mitigate them by integrating the GDPR legislation with the two security and privacy modelling techniques STRIDE and LINDDUN. The proposed technique focuses on \textit{(i)} proposing a new Data Flow Diagram for modellers to precisely describe a system taking into account GDPR-compliance, \textit{(ii)} building a knowledge base for GDPR-compliance, and \textit{(iii)} developing an inference engine for reasoning on the knowledge base. For the demonstration, we have applied the non-compliance threats in a telehealth service (i.e., non-consent, non-compliance with the Right to Erasure, and non-compliance with Accountability). The results have shown the feasibility and effectiveness of the modelling technique in addressing such threats.

For future work, we are developing a comprehensive modelling technique to cope with all non-compliance threats. This requires constructing a rule-based knowledge base for GDPR compliance. For this purpose, a GDPR-compliance ontology will be developed based on the methodology for building Legal Ontology (MeLOn) and integrated with other related ontologies. Another research direction is to develop a novel inference engine reasoning over a GDPR-compliance knowledge base. We plan to use Defeasible Logic, a non-monotonic formalism with conflicts-solving ability \cite{garcia2004defeasible}, that supports the legal reasoning and compliance checking.

\bibliography{refs}
\bibliographystyle{IEEEtran}

\end{document}